\documentclass[12pt]{article}
\usepackage{graphics}
\makeatletter

\renewcommand{\section}{\@startsection {section}{1}{\z@}
{-3.5ex plus -1ex minus -.2ex}{2.3ex plus .2ex}{\normalsize\bf}}

\renewcommand{\subsection}{\@startsection{subsection}{2}{\z@}
{-3.25ex plus -1ex minus -.2ex}{1.5ex plus .2ex}{\normalsize\it}}

\def\abstract{\if@twocolumn
\section*{\abstractname}
\else \small
\quotation
\fi}
\def\endabstract{\if@twocolumn\else\endquotation\fi}

\renewcommand{\@makefnmark}{\hbox{\mathsurround=0pt
$^\dagger$}}
\renewcommand{\@makefntext}[1]{\parindent=1em\noindent
\hbox to 1.8em{\hss$^\dagger$}#1}

\def\thebibliography#1{\section*{\refname\@mkboth
 {\uppercase{\refname}}{\uppercase{\refname}}}\list
 {[\arabic{enumi}]}{\settowidth\labelwidth{[#1]}\leftmargin\labelwidth
 \advance\leftmargin\labelsep\parsep=0em\itemsep=0em
 \usecounter{enumi}}
 \def\newblock{\hskip .11em plus .33em minus .07em}
 \sloppy\clubpenalty4000\widowpenalty4000
 \sfcode`\.=1000\relax}

\makeatother

\begin{document}

\begin{center}

{\normalsize\bf
Correlation Energies in Distorted $3d$-$t_{2g}$ Perovskite Oxides
}\\
\bigskip
I.~V.~Solovyev\footnote{
e-mail: solovyev.igor@nims.go.jp}
\medskip \\
{\small\it
Computational Materials Science Center (CMSC),\\
National Institute for Materials Science (NIMS),\\
1-2-1 Sengen, Tsukuba, Ibaraki 305-0047, Japan}

\end{center}

\begin{abstract}
Using an effective low-energy Hamiltonian derived
from the first-principles
electronic structure calculations for the narrow $t_{2g}$ bands
of YTiO$_3$, LaTiO$_3$, YVO$_3$, and LaVO$_3$,
we evaluate the contributions of the correlation energy ($E_C$)
to the stability of different magnetic structures, which can be realized in these
distorted perovskite oxides. We consider two approximations for $E_C$, which are based
on the regular perturbation theory expansion around a nondegenerate
Hartree-Fock ground state.
One is the second order of perturbation theory, which allows us to compare the
effects of
local and nonlocal correlations.
Another one is the local $t$-matrix approach, which allows us to treat some
higher-order contributions to $E_C$.
The correlation effects
systematically improve the agreement with the
experimental data and additionally stabilize the
experimentally observed $G$- and $C$-type antiferromagnetic (AFM) structures
in YVO$_3$ and LaVO$_3$, though the absolute magnitude of the stabilization energy
is sensitive to the level of
approximations and somewhat smaller in the $t$-matrix method.
The nonlocal correlations additionally stabilize the ferromagnetic ground state in
YTiO$_3$ and the $C$-type AFM ground state in LaVO$_3$.
Amongst two inequivalent transition-metal sites
in the monoclinic structure,
the local correlations
are stronger at
the sites with the least distorted environment.
Limitations of the regular perturbation-theory expansion for LaTiO$_3$ are also discussed.

\end{abstract}

\begin{sloppypar}

\section{Introduction}

  An interest
to the transition-metal perovskite oxides YTiO$_3$, LaTiO$_3$,
YVO$_3$, and LaVO$_3$ is mainly related with the variety of
magnetic structures, which can be realized in these, seemingly alike, compounds. For
example, YTiO$_3$ has the ferromagnetic ($F$) structure~\cite{YTiO3_exp}. LaTiO$_3$ is a
three-dimensional ($G$-type) antiferromagnet~\cite{LaTiO3_exp}. At the
low temperature, YVO$_3$ forms the $G$-type antiferromagnetic (AFM)
structure, which can be transformed to a chainlike ($C$-type)
antiferromagnetic structure at around $77$ K~\cite{YVO3_exp}. On the
contrary, LaVO$_3$ is the $C$-type antiferromagnet in the whole
temperature range below the magnetic transition
temperature~\cite{LaVO3_exp}. Surprisingly, the difference exists
not only between titanates (YTiO$_3$ and LaTiO$_3$) and vanadites
(YVO$_3$ and LaVO$_3$), which have a different number of valent
electrons, but also within each group of formally isoelectronic
materials. The differences are apparently related with the tiny
changes in the distorted perovskite
structure, which are amplified by the effects of Coulomb correlations
in the narrow $t_{2g}$ band. The details of the crystal structure
can be found in
Refs.~\cite{YTiO3_exp,LaTiO3_exp,YVO3_exp,LaVO3_exp}. Briefly, both
titanites have an orthorhombic structure, although the details of
this structure are rather different for YTiO$_3$ and LaTiO$_3$.
LaVO$_3$ is crystallized in a monoclinic structure. The
low-temperature phase of YVO$_3$ is orthorhombic (shown in
Fig.~\ref{fig.structure_DOS}), which becomes monoclinic at around
$77$ K. The structural orthorhombic-monoclinic transition
coincides with the $G$-$C$ AFM transition. Generally, Y-based compounds
are more distorted (due to
smaller size of the Y$^{3+}$ ions).

  There is a large number of theoretical articles devoted to the origin of the magnetic ground
states in the distorted $t_{2g}$ perovskite oxides.
The problem
has been considered on the basis of
first-principles electronic structure calculations
(e.g., Refs.~\cite{first_principles}) and the model approaches for the
strongly-correlated systems (e.g., Refs.~\cite{model1,model2.1,model2.2}).
The model theories typically vary on the assessment of the role played by the
lattice distortions~\cite{model1} and the Coulomb correlations~\cite{model2.1,model2.2}.

  We believe that any realistic theoretical description of these compounds
is practically impossible without the impact from the
first-principles
electronic structure calculations: simply, the lattice distortion is too
complex, and,
had we try to postulate a model Hamiltonian for these $t_{2g}$ perovskite oxides,
we would inevitably face the problem
of choosing the values for a large number of
model parameters, which cannot be fixed
in unbiased way. However, the conventional electronic structure calculations
are also far from being perfect. Typically, they are supplemented with some additional
approximations, which have serious limitations
for treating the Coulomb correlations
in the case of
strongly-correlated materials.
A typical example is the local-density approximation (LDA).
From this point of view, a promising direction is to make a bridge
between first-principles electronic structure calculations and models for the strongly-correlated
systems, and construct an appropriate model Hamiltonian entirely ``from the first principles''.
Fortunately, in the case of transition-metal oxides, we are typically dealing only with a
small group of states located near the Fermi level
and well separated from the remaining part of the spectrum
(for instance, $t_{2g}$ bands in Fig.~\ref{fig.structure_DOS}).
These states are mainly responsible for the electronic and magnetic properties of oxide materials.
Therefore, in may cases it is sufficient to consider a minimal model, consisting of
only the $t_{2g}$ bands,
and include the effect of other bands into the renormalization of interaction
parameters in the $t_{2g}$ band.
Such a strategy was pursued in Refs.~\cite{Imai,PRB06,condmat06}. It consists of
three major steps: first-principles electronic structure calculations $\rightarrow$
construction of the model Hamiltonian $\rightarrow$ solution of this model Hamiltonian.
The first applications to the distorted
$t_{2g}$ perovskite oxides have been considered in Refs.~\cite{condmat06,PRB04}.
The present paper deals with the last part of the problem. We will solve the
model Hamiltonian derived in Ref.~\cite{condmat06}, and mainly focus on the role
played by the correlation effects, beyond the mean-field Hartree-Fock (HF) approximation.
Particularly, we will consider two perturbative approaches.
One is the regular second-order perturbation theory for the
correlation energy~\cite{cor2ndorder}, and the other one is the $t$-matrix
approach~\cite{Brueckner,Galitskii,Kanamori}. In both approaches,
the HF approximation is used as the starting point. This
implies that the degeneracy
of the HF ground state is already lifted by the crystal distortion so that the regular
perturbation theory is justified. We will also discuss some limitations of this
treatment for LaTiO$_3$.

\section{\label{sec:construction}Construction of model Hamiltonian}

  Our first goal is the construction of the effective
multi-orbital Hubbard model for
the isolated $t_{2g}$ bands:
\begin{equation}
\hat{\cal{H}}= \sum_{{\bf R}{\bf R}'} \sum_{\alpha \beta} h_{{\bf
R}{\bf R}'}^{\alpha \beta}\hat{c}^\dagger_{{\bf
R}\alpha}\hat{c}^{\phantom{\dagger}}_{{\bf R}'\beta} + \frac{1}{2}
\sum_{\bf R}  \sum_{\alpha \beta \gamma \delta} U_{\alpha \beta
\gamma \delta} \hat{c}^\dagger_{{\bf R}\alpha} \hat{c}^\dagger_{{\bf
R}\gamma} \hat{c}^{\phantom{\dagger}}_{{\bf R}\beta}
\hat{c}^{\phantom{\dagger}}_{{\bf R}\delta},
\label{eqn:Hmanybody}
\end{equation}
where $\hat{c}^\dagger_{{\bf R}\alpha}$ ($\hat{c}_{{\bf R}\alpha}$)
creates (annihilates) an electron in the Wannier orbital
$\tilde{W}_{\bf R}^\alpha$ of the transition-metal site ${\bf R}$, and $\alpha$ is
a joint index, incorporating all remaining (spin and orbital)
degrees of freedom. The matrix $\hat{h}_{{\bf R}{\bf R}'}$$=
$$\| h_{{\bf R}{\bf R}'}^{\alpha \beta} \|$
parameterizes the kinetic energy of electrons, where
the site-diagonal part (${\bf R}$$=$${\bf R}'$) describes
the local level-splitting, caused by the crystal field, and the off-diagonal part
(${\bf R}$$\neq$${\bf R}'$) stands for the transfer integrals.
$U_{\alpha \beta \gamma \delta}
=
\int d{\bf r} \int d{\bf r}' \tilde{W}_{\bf R}^{\alpha \dagger}({\bf r})
\tilde{W}_{\bf R}^\beta({\bf r}) v_{\rm scr}({\bf r}$$-$${\bf r}')
\tilde{W}_{\bf R}^{\gamma \dagger}({\bf r}') \tilde{W}_{\bf R}^\delta({\bf r}')
\equiv \langle \alpha \gamma | v_{\rm scr} |
\beta \delta \rangle$
are
the matrix elements of \textit{screened} Coulomb interaction
$v_{\rm scr}({\bf r}$$-$${\bf r}')$, which are supposed to be diagonal with
respect to the site indices.
In principles, $U_{\alpha \beta \gamma \delta}$ can also depend on the
site-index ${\bf R}$. Nevertheless, for the sake of simplicity of
our notations, here and throughout in this paper we drop the index ${\bf R}$
in the notation of the Coulomb matrix elements (however, we do consider this
dependence in all our calculations).

  The procedure of mapping of the first-principles electronic structure
calculations onto the model Hamiltonian (\ref{eqn:Hmanybody})
for distorted
perovskite oxides has been discussed in details in Refs.~\cite{PRB06,condmat06}. Here,
we only outline the main idea. The kinetic-energy part,
$\hat{h}_{{\bf R}{\bf R}'}$, can be obtained using the downfolding method, which
is exact and equivalent to the projector-operator method~\cite{preprint06}.
The Wannier functions can be formally derived from $\hat{h}_{{\bf R}{\bf R}'}$,
using the ideas of the linear-muffin-tin-orbital (LMTO) method~\cite{PRB06,LMTO}.
The matrix of screened Coulomb interactions in the $t_{2g}$ band can be
calculated using a hybrid approach, which combines the constraint density-functional
theory with the random-phase approximation for the hybridization effects between
transition-metal $d$ and other atomic states~\cite{PRB06}.
The values of the model parameters obtained in such a way can be found in
Ref.~\cite{condmat06}.

\section{Solution of model Hamiltonian}

\subsection{Hartree-Fock approximation}

  The HF method provides the simplest approximation for the solution of the
many-electron
problem with the Hamiltonian (\ref{eqn:Hmanybody}).
In this case,
the trial many-electron
wavefunction is searched in the form of a single Slater determinant
$|S\{ \varphi_k \} \rangle$, constructed from the one-electron orbitals
$\{ \varphi_k \}$.
In these notations, $k$ is a joint index, which contains the information about
the momentum (${\bf k}$)
in the first Brillouin zone, the number of band ($n$), and the spin
($\sigma$$=$ $\uparrow$ or $\downarrow$) of the particle.
The one-electron orbitals
$\{ \varphi_k \}$ are subjected to the variational principle and requested to minimize
the total energy
$$
E_{\rm HF}= \min_{\{ \varphi_k \}}
\langle S\{ \varphi_k \}|\hat{\cal{H}}| S\{ \varphi_k \} \rangle
$$
for a given number of particles $\cal{N}$.
This yields the following equations for $\{ \varphi_k \}$:
\begin{equation}
\left( \hat{h}_{\bf k} + \hat{V} \right) | \varphi_k \rangle =
\varepsilon_k | \varphi_k \rangle,
\label{eqn:HFeq}
\end{equation}
where
$\hat{h}_{\bf k}$$\equiv$$\| h_{\bf k}^{\alpha \beta} \|$
is the kinetic part of the model Hamiltonian (\ref{eqn:Hmanybody}) in the reciprocal space:
$h_{\bf k}^{\alpha \beta}$$=
$$\frac{1}{N} \sum_{{\bf R}'} h^{\alpha \beta}_{{\bf R}{\bf R}'} e^{-i {\bf k} \cdot ({\bf R}-{\bf R}')}$
($N$ being the number of sites),
and $\hat{V}$$\equiv$$\| V_{\alpha \beta} \|$ is the HF potential:
\begin{equation}
V_{\alpha \beta} = \sum_{\gamma \delta}
\left( U_{\alpha \beta \gamma \delta} - U_{\alpha \delta \gamma \beta} \right)
n_{\gamma \delta}.
\label{eqn:HFpot}
\end{equation}
In the following, we will also use the notation $\hat{h}_{\rm HF}$,
which stands for the total Hamiltonian of the HF method,
$\hat{h}$$+$$\hat{V}$.
Eq.~(\ref{eqn:HFeq}) is solved self-consistently together with the equation
$$
\hat{n} = \sum_k^{occ} | \varphi_k \rangle \langle \varphi_k |
$$
for the
density matrix $\hat{n}$$\equiv$$\|n_{\alpha \beta}\|$ in the basis
of Wannier orbitals.
Finally,
the total energy in the HF method can be obtained as
$$
E_{\rm HF} = \sum_k^{occ} \varepsilon_k -\frac{1}{2}
\sum_{\alpha \beta} V_{\beta \alpha} n_{\alpha \beta}.
$$

\subsection{\label{sec:2ndorder}Second Order Perturbation Theory for Correlation Energy}

  The simplest way to go beyond the HF approximation is to include the correlation
interactions
in the second order of perturbation theory for the total energy~\cite{cor2ndorder}.
The correlation interaction (or a fluctuation)
is defined as the difference between true many-body
Hamiltonian~(\ref{eqn:Hmanybody}), and its one-electron
counterpart, obtained at the level of HF approximation:
\begin{equation}
\hat{\cal{H}}_C = \sum_{\bf R} \left(
\frac{1}{2} \sum_{\alpha \beta \gamma \delta}
U_{\alpha \beta \gamma \delta}
\hat{c}^\dagger_{{\bf R}\alpha} \hat{c}^\dagger_{{\bf R}\gamma}
\hat{c}^{\phantom{\dagger}}_{{\bf R}\beta} \hat{c}^{\phantom{\dagger}}_{{\bf R}\delta} -
\sum_{\alpha \beta} V_{\alpha \beta}
\hat{c}^\dagger_{{\bf R}\alpha} \hat{c}^{\phantom{\dagger}}_{{\bf R}\beta} \right).
\label{eqn:H2ndorder}
\end{equation}
By treating $\hat{\cal{H}}_C$ as a perturbation, the correlation energy can be easily
estimated as~\cite{cor2ndorder}:
\begin{equation}
E_C^{(2)} = - \sum_{S} \frac{
\langle G | \hat{\cal{H}}_C | S \rangle \langle S | \hat{\cal{H}}_C | G \rangle }
{E_{\rm HF}(S) - E_{\rm HF}(G)},
\label{eqn:dE2ndorder}
\end{equation}
where $|G \rangle$ and $|S \rangle$ are the Slater determinants
corresponding to the low-energy ground state
in the HF approximation, and the
excited state, respectively.
Due to the variational properties of the
HF method, the only processes which may contribute to
$E_C^{(2)}$ are the
two-particle excitations, for which each
$|S \rangle$ is obtained from $|G \rangle$ by replacing
two one-electron orbitals, say $\varphi_{k_1}$
and $\varphi_{k_2}$, from the occupied part of the spectrum
by two unoccupied orbitals, say $\varphi_{k_3}$ and $\varphi_{k_4}$.
Hence, using the notations of Sec.~\ref{sec:construction}, the matrix elements take the
following form:
\begin{equation}
\langle S | \hat{\cal{H}}_C | G \rangle =
\langle k_3 k_4 | v_{\rm scr} |
k_1 k_2 \rangle -
\langle k_3 k_4 | v_{\rm scr} |
k_2 k_1 \rangle.
\label{eqn:dHmelement}
\end{equation}
Then, we employ a common approximation
of noninteracting quasiparticles
and replace the denominator of Eq.~(\ref{eqn:dE2ndorder})
by the linear combination of HF eigenvalues:
$E_{\rm HF}(S)$$-$$E_{\rm HF}(G) \approx \varepsilon_{k_3}$$+$$\varepsilon_{k_4}$$-
$$\varepsilon_{k_1}$$-$$\varepsilon_{k_2}$~\cite{cor2ndorder}.
The matrix elements (\ref{eqn:dHmelement})
satisfy the following condition:
$\langle S | \hat{\cal{H}}_C | G \rangle$$\sim$$\frac{1}{N}
\sum_{\bf R} e^{i({\bf k}_3+{\bf k}_4-{\bf k}_1-{\bf k}_2) \cdot {\bf R}}$,
provided that
the screened Coulomb interactions are diagonal with respect to the site
indices.

  A good point of the second-order of the perturbation theory is that
it allows us to estimate relatively easily both on-site (${\bf R}$$=$$0$) and
intersite (${\bf R}$$\ne$$0$) elements of
this expansion. In the following, we will use this method in order to study
the relative role played by these effects in the stability of
different magnetic structures of the distorted perovskite oxides.
The ${\bf R}$$=$$0$ term corresponds to the commonly used
single-site approximation for the correlation interactions,
which becomes exact in the limit of infinite spacial dimensions~\cite{DMFT}.

\subsection{\label{sec:tmatrix}$t$-matrix approach}

  The basic idea of the $t$-matrix approach is to look at
the true many-electron system as a superposition of
independent two-electron subsystems, and to solve rigorously
the Schr\"{o}dinger equations for
each
of these subsystems~\cite{Brueckner,Galitskii,Kanamori}.
Hence, we consider the following two-electron Hamiltonian:
$$
\hat{H}(1,2) = \hat{h}_{\rm HF}(1) + \hat{h}_{\rm HF}(2) + \Delta \hat{v}(1,2),
$$
where
$\Delta \hat{v}(1,2)$$=$$\hat{v}_{\rm scr}(1,2)$$-$$\hat{V}(1)$$-$$\hat{V}(2)$,
$\hat{v}_{\rm scr}(1,2)$ is the screened (by other bands) Coulomb
interactions between electrons `1' and `2' in the $t_{2g}$ band,
and $\hat{h}_{\rm HF}$ ($\hat{V}$) is the one-electron Hamiltonian (potential)
in the HF approximation.
For the periodic system,
the Schr\"{o}dinger equation can be
written in the following form:
\begin{equation}
\hat{H} | \Psi_{k_1 k_2} \rangle = E_{k_1 k_2} | \Psi_{k_1 k_2} \rangle.
\label{eqn:2eSchroedinger}
\end{equation}
Any two-electron wavefunction $| \Psi_{k_1 k_2} \rangle$
can be expanded in the basis of (also two-electron)
Slater's determinants: $| k_1 k_2 \rangle$$=$$\frac{1}{\sqrt{2}}
\{ \varphi_{k_1}(1) \varphi_{k_2}(2)$$-$$\varphi_{k_2}(1) \varphi_{k_1}(2) \}$, etc.
Apart from a
normalization multiplier, this expansion has the following form~\cite{Kanamori}:
\begin{equation}
| \Psi_{k_1 k_2} \rangle = | k_1 k_2 \rangle + \sum_{| k_3 k_4 \rangle} \Gamma_{k_1 k_2}^{k_3 k_4} | k_3 k_4 \rangle.
\label{eqn:2eWF}
\end{equation}
Note that the summation goes only over \textit{nonequivalent} Slater's determinant $| k_3 k_4 \rangle$,
constructed from the one-electron orbitals $k_3$ and $k_4$.
For example, since
$| k_4 k_3 \rangle$$=$$-$$| k_3 k_4 \rangle$, the determinant
$| k_4 k_3 \rangle$
should be excluded from the sum (\ref{eqn:2eWF}), etc.
By substituting Eq.~(\ref{eqn:2eWF}) into Eq.~(\ref{eqn:2eSchroedinger}), and
introducing the new notations
$\Delta E_{k_1 k_2}$$=$$E_{k_1 k_2}$$-$$\varepsilon_{k_1}$$-$$\varepsilon_{k_2}$,
such that
$$
\left[ \hat{h}_{\rm HF}(1) + \hat{h}_{\rm HF}(2)
- \varepsilon_{k_1} - \varepsilon_{k_2}
\right]  | k_1 k_2 \rangle  = 0
$$
(i.e., $\varepsilon_{k_1}$ and $\varepsilon_{k_2}$ are the eigenvalues of the HF Hamiltonian),
one obtains the following equation for $\Delta E_{k_1 k_2}$ and
$\Gamma_{k_1 k_2}^{k_3 k_4}$:
$$
\left( \Delta \hat{v} - \Delta E_{k_1 k_2} \right)  | k_1 k_2 \rangle +
\sum_{| k_3 k_4 \rangle} \left( \varepsilon_{k_3} + \varepsilon_{k_4} -
\varepsilon_{k_1} - \varepsilon_{k_2} + \Delta \hat{v} - \Delta E_{k_1 k_2} \right)
\Gamma_{k_1 k_2}^{k_3 k_4}  | k_3 k_4 \rangle = 0.
$$
By considering the matrix element of this equation with $\langle k_1 k_2 |$, one can find that
\begin{equation}
\Delta E_{k_1 k_2} = \langle k_1 k_2 | \Delta \hat{v} | k_1 k_2 \rangle +
\sum_{| k_3 k_4 \rangle} \Gamma_{k_1 k_2}^{k_3 k_4} \langle k_1 k_2 | \Delta \hat{v} | k_3 k_4 \rangle,
\label{eqn:2eEnergy}
\end{equation}
where the first term is the energy of Coulomb and exchange interactions
in the HF approximation
(minus the potential energy), while
the second term is the correlation energy.
By considering the matrix elements with $\langle k_5 k_6 |$$\ne$$\langle k_1 k_2 |$,
one can find another set of equations for $\Gamma_{k_1 k_2}^{k_3 k_4}$:
$$
\langle k_5 k_6 | \Delta \hat{v} | k_1 k_2 \rangle +
\left( \varepsilon_{k_5} + \varepsilon_{k_6} - \varepsilon_{k_1} - \varepsilon_{k_2}
- \Delta E_{k_1 k_2} \right) \Gamma_{k_1 k_2}^{k_5 k_6} +
\sum_{| k_3 k_4 \rangle} \Gamma_{k_1 k_2}^{k_3 k_4} \langle k_5 k_6 | \Delta \hat{v} | k_3 k_4 \rangle
= 0.
$$
They are solved iteratively with respect to $\Delta \hat{v}$.
In order to do so, it is convenient to introduce the two-particle Green's function,
$$
\hat{G}_{k_1 k_2} = \sum_{| k_3 k_4 \rangle} \frac{ | k_3 k_4 \rangle \langle k_3 k_4 | }
{\varepsilon_{k_3} + \varepsilon_{k_4} -
\varepsilon_{k_1} - \varepsilon_{k_2} - \Delta E_{k_1 k_2}},
$$
and derive a matrix equation for $\{ \Gamma_{k_1 k_2}^{k_3 k_4} \}$, which are then substituted
into Eq.~(\ref{eqn:2eEnergy}).
Then, it is rather straightforward to derive the following expression for $\Delta E_{k_1 k_2}$:
\begin{equation}
\Delta E_{k_1 k_2} = \langle k_1 k_2 | \hat{T}_{k_1 k_2} | k_1 k_2  \rangle,
\label{eqn:tmatrix_energy}
\end{equation}
where $\hat{T}_{k_1 k_2}$ is the so-called $t$-matrix:
\begin{equation}
\hat{T}_{k_1 k_2} = \Delta \hat{v} \left[ \hat{1} + \hat{G}_{k_1 k_2} \Delta \hat{v} \right]^{-1}.
\label{eqn:tmatrix}
\end{equation}
The correlation energy of the $t$-matrix method is obtained after the subtraction from
Eq.~(\ref{eqn:tmatrix_energy}) the energies
of Coulomb and exchange interactions in the HF approximation
and summation up over all Slater's
determinants constructed from the occupied one-electron orbitals of the HF method:
\begin{equation}
E_C^{(t)} = \sum_{| k_1 k_2 \rangle}^{occ} \langle k_1 k_2 | \hat{T}_{k_1 k_2} - \Delta \hat{v} | k_1 k_2  \rangle.
\label{eqn:tmatrix_correlation_energy}
\end{equation}
In practice,
each HF orbital has been expanded over the basis of Wannier functions, and
then all calculations of $\hat{T}_{k_1 k_2}$ and $E_C^{(t)}$ have been
performed in this basis.

  By expanding $\hat{T}_{k_1 k_2}$ up to the second order
of $\Delta \hat{v}$, we regain Eq.~(\ref{eqn:dE2ndorder}),
obtained
in the second order of perturbation theory.
Therefore, the good point of the $t$-matrix approach is that it allows us
to go beyond the second order of perturbation theory and evaluate the
higher order effects of $\Delta \hat{v}$ onto the
correlations energy. Nevertheless,
it was supplemented with some additional approximations.
\begin{enumerate}
\item
When we compute the matrix elements of the form
$\langle k_3 k_4 | \Delta \hat{v} | k_1 k_2 \rangle$,
being proportional to
$\frac{1}{N}
\sum_{\bf R} e^{i({\bf k}_3+{\bf k}_4-{\bf k}_1-{\bf k}_2) \cdot {\bf R}}$,
we consider only the ${\bf R}$$=$$0$ part of this sum and neglect all other
contributions.
This corresponds to the single-site approximation for the $t$-matrix.
\item
In all matrix elements $\langle k_3 k_4 | \Delta \hat{v} | k_1 k_2 \rangle$, we
replace $\Delta \hat{v}$ by $\hat{v}_{\rm scr}$ and
drop the one-electron potentials of the HF method.
Strictly speaking, this procedure is justified only when both
one-electron
states $k_1$ and $k_2$ are different from $k_3$ and $k_4$, for example,
when they belong, correspondingly, to the occupied and unoccupied part
of the spectrum, like in the second order of perturbation theory.
However, this is no longer true for the higher-order terms with respect to
$\Delta \hat{v}$. Nevertheless, we believe that the difference is small.
\end{enumerate}

  All correlation energies have been computed in the mesh of 75
points in the first Brillouin zone (BZ), corresponding to the 4:4:2
divisions of the reciprocal translation vectors for the distorted
perovskite structure.
The actual integration over the BZ has been replaced by the summation over this
mesh of points.

\section{Results and Discussions}

  First applications of the proposed method to YTiO$_3$, LaTiO$_3$,
YVO$_3$, and LaVO$_3$ have been considered in Ref.~\cite{condmat06},
where we have summarized results of HF calculations for the model
(\ref{eqn:Hmanybody}) and the behavior of correlation energies
in the second order of perturbation theory, supplemented with the
single-site approximation. In the present work we will
further elaborate the problem by
focusing on the following question:
\begin{enumerate}
\item
the role of
higher-order contributions to the correlation energy;
\item
the role of
nonlocal (or intersite) contributions to the correlation energy.
\end{enumerate}
We will also consider the effects of monoclinic distortion and
analyze the
contributions to the correlation energy of inequivalent transition-metal sites.
The results of these calculations are presented in Tables~\ref{tab:YTiO3}-\ref{tab:LaTiO3}
for all considered compounds.
First, we would like to summarize the main results of Ref.~\cite{condmat06}.
\begin{enumerate}
\item
The HF approximation yields
the correct magnetic ground state for YTiO$_3$, LaVO$_3$, and
both phases of YVO$_3$.
This conclusion is fully
consistent with the results of
accurate
all-electron band-structure calculations~\cite{first_principles},
and
this is quite remarkable that all these results can be reproduced in our minimal
model derived for the $t_{2g}$ bands.
\item
The correlation effects
favor the AFM spin alignment and
additionally stabilize the experimentally
observed $G$- and $C$-type AFM states in YVO$_3$ and LaVO$_3$.
\item
None of the considered approaches reproduces the experimental $G$-type AFM ground state
of LaTiO$_3$ (instead, the theoretical calculations steadily converge to the
$A$-type AFM ground state~\cite{condmat06,PRB04}).
\end{enumerate}
Then, what will happen if we go beyond the second-order perturbation theory and
apply the $t$-matrix approach? Generally, the $t$-matrix approach reduces the absolute value
of the correlation energy.
However, the magnitude of this reduction strongly depends on the magnetic state.
For example, if the $F$ state is only weakly affected by the higher-order
correlation effects (the typical changes of $E_C$ varies from 1\% in YVO$_3$ till 13\% in LaTiO$_3$),
$E_C$
in the $G$-type AFM phase can drop by nearly 50\%.
From this point of view, if the second order or perturbation theory does not solve
the problem of the $G$-type AFM ground state of LaTiO$_3$, it seems to be unlikely that
the higher-order effects can reverse the situation.
Apparently, LaTiO$_3$
is different from
other
perovskite oxides,
and
the regular
perturbation-theory expansion,
though may be justified for the majority of considered compounds,
does not work in the case of LaTiO$_3$.
This seems to be reasonable, because LaTiO$_3$ has the largest correlation energies,
which are comparable with the splitting of the $t_{2g}$ levels caused by the crystal
distortion ($\sim$$37$ meV~\cite{condmat06}).
Therefore, it is quite possible that the correlation effects in LaTiO$_3$ should be
considered \textit{at the first place}, and the simple HF theory for the spin and
orbital ordering with the subsequent inclusion of the correlation effects as a
perturbation to the HF ground state may not be appropriate here~\cite{model2.1,model2.2}.
Note that in other materials, the situation is different: the typical values
of the $t_{2g}$-levels splitting in YTiO$_3$, YVO$_3$, and LaVO$_3$ are about $100$ meV~\cite{condmat06},
which exceeds the correlation energy by at least one order of magnitude.
Therefore, it seems that the degeneracy of the HF ground state is already lifted
by the crystal distortion, and the correlation effects are well described by
means of the
regular perturbation theory expansion.
This is partly supported by recent total energy
calculations for the orthorhombic phase of YVO$_3$
using
path-integral renormalization group method,
which is free of
any perturbation-theory expansions for the
correlation energy~\cite{OtsukaImada}.
The method was applied to the same model, and the main conclusions concerning the
magnetic phase diagram were
similar to our present finding.

  The correlations additionally stabilize the experimentally observed
$G$- and $C$-type AFM states in YVO$_3$ and LaVO$_3$.
Moreover, in the orthorhombic phase of YVO$_3$, the correlation effects tend to stabilize the
$G$-type AFM state; while in the monoclinic phase, they stabilize the
$C$-type AFM state, being in total agreement with the experimental data.
This trend is clearly seen
both in the second order of perturbation theory and in the $t$-matrix approach,
though
the latter yield somewhat smaller values for
the stabilization energy associated with the correlation effects.

  The higher-order correlations play an important role YTiO$_3$ and additionally
stabilize the ferromagnetic phase. The latter emerges as the ground state already
in the HF approach, where the total energy difference between ferromagnetic and the
next $A$-type AFM state is about $2.05$ meV per one Ti site (Table~\ref{tab:YTiO3}).
However, if we take into account the correlation effects in the second order
of perturbation theory (and consider the single-site approximation),
this difference is reduced to only $0.99$ meV. Therefore, the situation is very
fragile.
Nevertheless,
the $t$-matrix approach, which affects
more strongly the $A$-type AFM state, will recover some
of these energy gains and make the total energy difference between
ferromagnetic and $A$-type AFM states
to be
about $1.83$ meV per one Ti site.

\begin{figure}
\begin{center}
\resizebox{14cm}{!}{\includegraphics{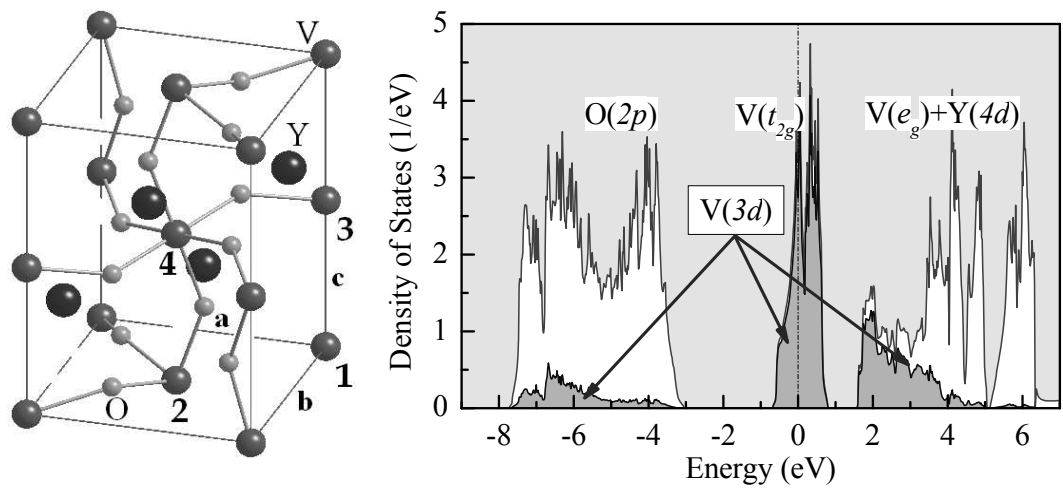}}
\end{center}
  \caption{\small
A characteristic example of the crystal structure (left) and the electronic
structure in the local-density approximation (right)
of the orthorhombically distorted YVO$_3$.
In the left panel,
the symbols ${\bf a}$, ${\bf b}$, and ${\bf c}$ stand for
orthorhombic translations, and the symbols 1--4 denote the transition-metal sites,
which form the unit cell of the distorted perovskite oxides.
In the right panel,
the shaded area shows contributions of the atomic V($3d$) states.
Other symbols show the positions of the main bands. The Fermi level is
at zero energy.
}
  \label{fig.structure_DOS}
\end{figure}

  The intersite correlation energies, which have been estimated
in the second order of perturbation theory, can be large in some
ferromagnetically couple bonds. This is especially true for YTiO$_3$
and LaVO$_3$. For example, the energy of interaction between nearest-neighbor
sites `1' and `2' (see Fig.~\ref{fig.structure_DOS}),
located in the
${\bf ab}$-plane of the ferromagnetic phase of YTiO$_3$ is about $-$$0.38$ meV
(Table~\ref{tab:YTiO3}).
Since in the
${\bf ab}$-plane,
each transition-metal atom interacts with four nearest neighbors,
it corresponds to the additional energy gain
$-$$0.38$$\times$$4=-$$1.52$ meV per one Ti site.
Similar estimates yield
$-$$0.37$$\times$$4=-$$1.48$ meV,
$-$$0.17$$\times$$4=-$$0.68$ meV, and
$-$$0.18$$\times$$4=-$$0.72$ meV, correspondingly for the $A$-, $C$-, and $G$-type
AFM states.
Therefore, the in-plane intersite correlations tend to additionally
stabilize the ferromagnetic phase relative to the
AFM states
$C$ and $G$.
In the $A$-type AFM phase,
the sites `1'and `2' are also ferromagnetically coupled,
like in the totally ferromagnetic phase.
Therefore,
these two phases have practically the same
intersite correlation energies in the ${\bf ab}$-plane.
The inter-plane correlations appear to be small in all magnetic phases of YTiO$_3$.

  In LaVO$_3$, the situation is somewhat different, and this is a good
example of the system where already the inter-plane correlations play a
more important role.
Indeed, the energies of intersite correlations are the largest
in the ferromagnetic chains of $C$-type AFM phase, which is also the
magnetic ground state of this compound.
These energies are associated with the bonds `1-3' and `2-4', which are shown in Fig.~\ref{fig.structure_DOS},
and the results are summarized in Table~\ref{tab:LaVO3}.
Thus, in the case of LaVO$_3$, the inter-plane correlations additionally stabilize the
$C$-type AFM ground state. However, since each transition-metal atom interacts with only
two nearest neighbors along the ${\bf c}$ axis, the stabilization energy is not
particularly large
(about $-$$0.26$$\times$$2=-$$0.52$ meV per one V atom).

  The intersite correlation energies are large also in the case of LaTiO$_3$ (Table~\ref{tab:LaTiO3}).
However, they tend to stabilize either ferromagnetic or $A$-type AFM states, and do not
explain
the appearance of experimental $G$-type AFM ground state. Again, we believe that the problem is related
with the use of the regular perturbation theory expansion,
which may not be justified in the case of LaTiO$_3$.

  The monoclinic distortion realized in LaVO$_3$ and in the high-temperature phase
of YVO$_3$ produces two inequivalent pairs of transition-metal sites,
which are shown correspondingly as
(1,2) and (3,4) in Fig.~\ref{fig.structure_DOS}.
Therefore, it is interesting to consider the interplay between correlation energies
and the lattice distortions around
different transition-metal sites.
In our notations, the crystal structure
around the sites `3' and `4' is more distorted than the one
around the
sites `1' and `2'.
Such a distortion directly correlates with the
magnitude of the crystal-field splitting in different sublattices~\cite{condmat06}.
Then, the
on-site correlations are generally stronger at the sites with the
least distorted environment (site `1' in the Tables~\ref{tab:YVO3m} and~\ref{tab:LaVO3}).
This rule holds both for YVO$_3$ and LaVO$_3$ (though with some exception for the
ferromagnetic phase of LaVO$_3$).
In the $C$-type AFM phase,
which is always realized as the magnetic ground state in the monoclinic structure,
the
difference of on-site correlation energies associated with the sites `1' and `4'
is about $1$ meV per one V site, as obtained
in the second order of perturbation theory.
This value is further reduced
till $0.5$ meV per one V site by
higher-order correlations in the $t$-matrix theory.

\section{Summary and Conclusions}

  This paper is the continuation of previous works
(Refs.~\cite{PRB06}, \cite{condmat06}, and \cite{PRB04})
devoted to the
construction and solution of an effective low-energy models for the series of distorted
$t_{2g}$ perovskite oxides on the basis of first-principles electronic structure
calculations.
It deals with the analysis of correlation interactions
and their contributions to stability
on different magnetic structures, which can be realized in these compounds.
The correlation energies have been calculated on the basis of a regular
perturbation theory expansion starting from the ground state of the HF method.
Thus, our strategy implies that the degeneracy of the HF ground state is already lifted by
the crystal distortion and the regular perturbation theory is justified.
This seems to be a good approximation for the most distorted YTiO$_3$, YVO$_3$, and
even LaVO$_3$, where
\begin{enumerate}
\item
the correct magnetic ground state can be formally obtained at the level of HF approximation;
\item
the correlation effects, included as a
perturbation to the HF ground state, systematically improve the agreement with the experimental
data.
\end{enumerate}
However, in LaTiO$_3$, the situation is completely different:
\begin{enumerate}
\item
the HF method yields an incorrect
magnetic ground state ($A$-type AFM instead of $G$-type AFM);
\item
the correlation interactions, treated as a perturbation to this incorrect HF ground state,
do not change the overall picture, and the $G$-type AFM state remains
unstable relative to the $A$ state.
\end{enumerate}
Thus, the origin of the $G$-type AFM ground state in LaTiO$_3$
seems to be different from other perovskite oxides and remains a challenging problem for
future theories.
Apparently,
one of our basic assumptions about the nondegeneracy of the HF ground state
breaks down in the case of LaTiO$_3$, and
the true ground state
cannot be approached
through the series of continuous corrections applied
to the
single-Slater-determinant HF theory.
Therefore, the next important step for LaTiO$_3$ would be get rid of this
``nondegeneracy assumption''
and expand the class of the possible
ground states,
which would include some aspects of
the orbital liquid theory~\cite{model2.1}.

\section*{ACKNOWLEDGMENTS}
This work has been partially supported by Grant-in-Aids
for Scientific Research in Priority Area ``Anomalous Quantum Materials''
from the Ministry of Education, Culture, Sports, Science and Technology of Japan.

\end{sloppypar}

\newpage

\small{

}  

\newpage

\begin{table}
\caption{Hartree-Fock, $E_{\rm HF}$, and correlation energies obtained in the
second order of perturbation theory, $E_C^{(2)}$, and in the
$t$-matrix approach, $E_C^{(t)}$, for the orthorhombic phase of YTiO$_3$.
The Hartree-Fock energies are measured from the most stable magnetic state
in meV per one formula unit.
The correlation energies are measured in meV per one transition-metal site or a
pair of sites, correspondingly for the on-site and intersite contributions.
Note that the $t$-matrix was computed in the single-site approximation.
Therefore, only the site-diagonal part of $E_C^{(t)}$ is shown.
The positions of the transition-metal sites are shown in
Fig.~\protect\ref{fig.structure_DOS}.}
\label{tab:YTiO3}
\begin{center}
\begin{tabular}{cccccc}
\hline \hline
Phase     & $E_{\rm HF}$      & \multicolumn{3}{c}{$E_C^{(2)}$} & $E_C^{(t)}$  \\
\cline{3-6}
          &                   & Ti$_1$    & Ti$_1$-Ti$_2$   &  Ti$_1$-Ti$_3$     & Ti$_1$                        \\
\hline
   $F$    & $0$               & $-$$5.13$ & $-$$0.38$ & $-$$0.01$                &  $-$$4.58$                    \\
   $A$    & $\phantom{1}2.05$ & $-$$6.19$ & $-$$0.37$ &  $0$                     &  $-$$4.80$                    \\
   $C$    & $14.40$           & $-$$8.32$ & $-$$0.17$ & $-$$0.01$                &  $-$$5.28$                    \\
   $G$    & $16.25$           & $-$$8.48$ & $-$$0.18$ & $-$$0.01$                &  $-$$5.31$                    \\
\hline \hline
\end{tabular}
\end{center}
\end{table}

\

\newpage

\begin{table}
\caption{
Hartree-Fock, $E_{\rm HF}$, and correlation energies obtained in the
second order of perturbation theory, $E_C^{(2)}$, and in the
$t$-matrix approach, $E_C^{(t)}$, for the
low-temperature orthorhombic phase of YVO$_3$ ($T$$<$ $77$ K).
All energies are measured
in meV.
See Fig.~\protect\ref{tab:YTiO3} for the details of the notations.
}
\label{tab:YVO3o}
\begin{center}
\begin{tabular}{cccccc}
\hline \hline
Phase     & $E_{\rm HF}$ & \multicolumn{3}{c}{$E_C^{(2)}$} & $E_C^{(t)}$   \\
\cline{3-6}
          &              & V$_1$     & V$_1$-V$_2$ & V$_1$-V$_3$ &     V$_1$               \\
\hline
   $F$    & $21.66$      & $-$$2.19$ & $-$$0.12$ & $-$$0.02$ & $-$$2.16$           \\
   $A$    & $14.59$      & $-$$4.67$ & $-$$0.12$ & $-$$0.01$ & $-$$3.31$           \\
   $C$    & $10.14$      & $-$$5.61$ & $-$$0.07$ & $0$       & $-$$3.14$           \\
   $G$    & $0$          & $-$$7.07$ & $-$$0.07$ & $-$$0.01$ & $-$$4.06$           \\
\hline \hline
\end{tabular}
\end{center}
\end{table}

\

\newpage

\begin{table}
\caption{
Hartree-Fock, $E_{\rm HF}$, and correlation energies obtained in the
second order of perturbation theory, $E_C^{(2)}$, and in the
$t$-matrix approach, $E_C^{(t)}$, for the
high-temperature monoclinic phase of YVO$_3$ ($77$ K $<$$T$$<$ $116$ K).
All energies are measured
in meV.
See
Table~\protect\ref{tab:YTiO3} for the details of the notations.
Note that, in the monoclinic phase, the planes 1-2 and 3-4
are inequivalent (see Fig.~\protect\ref{fig.structure_DOS}).
Therefore, there are two different types of on-site (denoted as V$_1$ and V$_4$) and
intersite (denoted as V$_1$-V$_2$ and V$_4$-V$_3$) contributions to the
correlation energy. The contributions V$_1$-V$_3$ and V$_4$-V$_2$
are equivalent and are both shown only for the sake of completeness.
}
\label{tab:YVO3m}
\begin{center}
\begin{tabular}{cccccccccc}
\hline \hline
Phase     & $E_{\rm HF}$ & \multicolumn{6}{c}{$E_C^{(2)}$} & \multicolumn{2}{c}{$E_C^{(t)}$} \\
\cline{3-10}
          &                   & V$_1$     & V$_1$-V$_2$ & V$_1$-V$_3$ & V$_4$ & V$_4$-V$_3$ & V$_4$-V$_2$ & V$_1$ & V$_4$ \\
\hline
   $F$    & $11.71$           & $-$$2.81$ & $-$$0.02$ & $-$$0.03$ & $-$$1.74$ & $-$$0.01$ & $-$$0.03$ & $-$$2.76$ & $-$$1.71$            \\
   $A$    & $13.97$           & $-$$5.87$ & $-$$0.03$ & $-$$0.01$ & $-$$3.63$ & $-$$0.01$ & $-$$0.01$ & $-$$4.14$ & $-$$2.55$            \\
   $C$    & $0$               & $-$$8.08$ & $-$$0.02$ & $-$$0.05$ & $-$$6.98$ & $-$$0.03$ & $-$$0.05$ & $-$$4.85$ & $-$$4.33$            \\
   $G$    & $\phantom{1}6.63$ & $-$$7.56$ & $-$$0.02$ & $-$$0.01$ & $-$$6.49$ & $-$$0.03$ & $-$$0.01$ & $-$$4.38$ & $-$$3.76$            \\
\hline \hline
\end{tabular}
\end{center}
\end{table}

\

\newpage

\begin{table}
\caption{
Hartree-Fock, $E_{\rm HF}$, and correlation energies obtained in the
second order of perturbation theory, $E_C^{(2)}$, and in the
$t$-matrix approach, $E_C^{(t)}$, for the
monoclinic phase of LaVO$_3$.
All energies are measured in meV.
See
Tables~\protect\ref{tab:YTiO3} and \protect\ref{tab:YVO3m} for the details
of the notations.
Note that, in the monoclinic phase, the planes 1-2 and 3-4 (see Fig.~\protect\ref{fig.structure_DOS})
are inequivalent, that results in two types of V sites as well as the in-plane interactions.
}
\label{tab:LaVO3}
\begin{center}
\begin{tabular}{cccccccccc}
\hline \hline
Phase     & $E_{\rm HF}$ & \multicolumn{6}{c}{$E_C^{(2)}$} & \multicolumn{2}{c}{$E_C^{(t)}$} \\
\cline{3-10}
          &                   & V$_1$       & V$_1$-V$_2$ & V$_1$-V$_3$ & V$_4$ & V$_4$-V$_3$ & V$_4$-V$_2$ & V$_1$     & V$_4$   \\
\hline
   $F$    & $20.98$           & $-$$\phantom{1}3.82$ & $-$$0.02$ & $-$$0.15$ & $-$$\phantom{1}4.13$   & $-$$0.02$ & $-$$0.15$ & $-$$3.74$ & $-$$4.02$            \\
   $A$    & $20.63$           & $-$$11.77$           & $-$$0.22$ & $-$$0.03$ & $-$$\phantom{1}8.80$   & $-$$0.02$ & $-$$0.03$ & $-$$8.34$ & $-$$5.84$            \\
   $C$    & $0$               & $-$$13.37$           & $-$$0.04$ & $-$$0.26$ & $-$$12.54$             & $-$$0.02$ & $-$$0.26$ & $-$$8.86$ & $-$$8.39$            \\
   $G$    & $\phantom{1}7.65$ & $-$$10.52$           & $-$$0.04$ & $-$$0.02$ & $-$$\phantom{1}9.02$   & $-$$0.03$ & $-$$0.02$ & $-$$6.17$ & $-$$5.41$            \\
\hline \hline
\end{tabular}
\end{center}
\end{table}

\

\newpage

\begin{table}
\caption{
Hartree-Fock, $E_{\rm HF}$, and correlation energies obtained in the
second order of perturbation theory, $E_C^{(2)}$, and in the
$t$-matrix approach, $E_C^{(t)}$, for the
orthorhombic phase of LaTiO$_3$.
All energies are measured
in meV.
See
Table~\protect\ref{tab:YTiO3} for the details of the notations.
}
\label{tab:LaTiO3}
\begin{center}
\begin{tabular}{cccccc}
\hline \hline
Phase     & $E_{\rm HF}$      & \multicolumn{3}{c}{$E_C^{(2)}$} & $E_C^{(t)}$   \\
\cline{3-6}
          &                   & Ti$_1$     & Ti$_1$-Ti$_2$ & Ti$_1$-Ti$_3$ & Ti$_1$             \\
\hline
   $F$    & $\phantom{1}4.95$ & $-$$11.08$ & $-$$0.52$ & $-$$0.08$ & $-$$\phantom{1}9.66$       \\
   $A$    & $0$               & $-$$22.53$ & $-$$0.54$ & $-$$0.07$ & $-$$15.17$                 \\
   $C$    & $19.57$           & $-$$17.19$ & $-$$0.23$ & $-$$0.11$ & $-$$11.04$                 \\
   $G$    & $11.51$           & $-$$23.02$ & $-$$0.22$ & $-$$0.09$ & $-$$13.99$                 \\
\hline \hline
\end{tabular}
\end{center}
\end{table}

\

\newpage

\end{document}